\documentclass[a4paper,12pt]{article}
\usepackage{cite}

\advance\textwidth 0.5in \advance\oddsidemargin -0.25in \advance\textheight 0.5in \advance\topmargin -0.5in

\def\al{\alpha} \def\be{\beta} \def\ga{\gamma} \def\de{\delta}

\def\De{\Delta}  

\def\ph{\varphi}  

\def\beea{\begin{eqnarray}}
\def\eea{\end{eqnarray}}

\def\pa{\partial} \def\ti{\tilde}  \def\({\left(} \def\){\right)}

\def\beq{\begin{equation}} \def\ee{\end{equation}} \def\bea{\begin{array}} \def\ea{\end{array}}

\newtheorem{The}{Theorem}
\newtheorem{prop}{Proposition}
\def\proof{\paragraph{Proof.}} \def\qed{\vrule height0.6em width0.3em depth0pt}

\font\Sets=msbm10 \def\Zz{\hbox{\Sets Z}}

\title {\bf The Generalized Symmetry Method for Discrete Equations}

\author{{\bf D. Levi} \\ Dipartimento di Ingegneria Elettronica, \\ Universit\`a degli Studi Roma Tre and Sezione INFN, Roma Tre, \\ Via della
Vasca Navale 84, 00146 Roma, Italy \\ {\sl E-mail: levi@roma3.infn.it} \and {\bf R.I. Yamilov} \\ Ufa Institute of Mathematics, Russian Academy
of Sciences, \\ 112 Chernyshevsky Street, Ufa 450077, Russian Federation \\ {\sl E-mail: RvlYamilov@matem.anrb.ru}}

\date{\today} 
\begin{document} \maketitle

\begin{abstract} The generalized symmetry method is applied to a class of completely discrete equations including the Adler-Bobenko-Suris list.
Assuming the existence of a generalized symmetry, we derive a few integrability conditions suitable for testing and classifying equations of this
class. Those conditions are used at the end to test for integrability discretizations of some well-known hyperbolic equations. \end{abstract}


\section{Introduction}\label{s1}

The discovery of new two-dimensional integrable partial difference equations
(or $\Zz^2$-lattice equations)
 is always a very challenging problem as, by proper continuous
limits, many other results on integrable
differential-difference and partial differential equations may be obtained.  

The basic theory  and results in nonlinear
integrable differential  equations  can be found, for example,  in the
{\it Encyclopedia of Mathematical Physics} \cite{emp} or in the {\it Encyclopedia of Nonlinear Science} \cite{ens}.

The classification
of integrable nonlinear partial differential equations has been widely discussed in many
relevant papers. Let us just
mention here the classification scheme  introduced by Shabat, where the formal symmetry
approach has been introduced (see \cite{msy,MSS} for a review). This approach has been successfully extended to the differential-difference case
by Yamilov \cite{y83,y06,ly,asy}. In the completely discrete case the situation
turns out to be quite different, and, up to now, 
the formal symmetry technique has not been able to provide any result. In the case of difference--difference equations, the first
exhaustive classification
has been obtained in \cite{a2000} by Adler and in \cite{abs,abs1,asu,bob} by Adler, Bobenko and Suris for linear affine equations. The so obtained equations have been thoroughly studied by many researchers, and it has been shown that they have Lax pairs and that possess generalized symmetries \cite{RH,LPSY,gr,gr1,ni}. 

We study in this paper the following class of autonomous discrete equations on the lattice $\Zz^2$:
\beq\label{i1} u_{i+1,j+1} = F(u_{i+1,j}, u_{i,j}, u_{i,j+1}) ,
\ee
where $i,j$ are arbitrary integers. Many integrable examples of equations of this form are known \cite{abs,lp,lps,Q5,gr}. There are a number of papers, in which various schemes for testing and classifying integrable equations of the form (\ref{i1}) are discussed \cite{AS,H,Viallet,Ramani,Ablowitz_Hal,hal_ke,hal,abs,SRH}. In \cite{AS,H} the classification of linearizable equations is considered, in \cite{Viallet,Ramani,Ablowitz_Hal,hal_ke,hal} extensions of the Painlev\'e test are carried over to the discrete case while in \cite{abs,SRH} equations integrable by the inverse scattering method are discussed. Requiring additional geometrical symmetry properties, a classification result has been obtained in \cite{abs} together with a list of integrable equations. However, the symmetries for those discrete equations obtained in \cite{RH,gr} show that the obtained class of equations contained in \cite{abs} is somehow restricted \cite{LPSY}. From \cite{LPSY} it follows that one should expect a larger number of integrable discrete equations of the kind of eq. (\ref{i1}) than those up to now known.

Eqs. (\ref{i1}) are possible discrete analogs of the hyperbolic equations
  \beq\label{i2} u_{x,y} = F(u_x, u, u_y) . \ee
Eqs. (\ref{i2}) are very important in many fields of physics, and, as such,  they have been studied using the generalized symmetry method, however without much success. Only  the following two particular cases: \beq\label{i3} u_{x,y} = F(u) ,
\ee \beq\label{i4} u_x = F(u,v) , \quad v_y = G(u,v) , \ee which are essentially easier, have been solved \cite{ZS79,ZS84}. The study of the class of equations (\ref{i1}) may be important to characterize the integrable subcases of eq. (\ref{i2}).

In Section \ref{s2}  we introduce and discuss some necessary notions, such as  generalized symmetries and conservation laws for discrete systems of the form (\ref{i1}), and in Section \ref{s3a}, as a motivation for the use of this approach, we show that one can construct a partial difference equation closely related to the modified Volterra equation, which does not belong to the ABS class of equations as it is not 3D--consistent around the cube and does not have the $D_4$ symmetry. In Section \ref{s3}, following the standard scheme of the generalized symmetry method, we derive a few integrability conditions for the class (\ref{i1}). These conditions are not sufficient to carry out a classification of the discrete equations (\ref{i1}). So in Section \ref{s4} we reduce ourselves to consider just 5 points generalized symmetries. This request provides further integrability conditions. With these extra conditions, the set of obtained conditions  will be suitable for testing and classifying  simple classes of difference equations of the form (\ref{i1}). As an example, in Section \ref{s5} we apply those conditions to the class of equations  
 \beq\label{i5} u_{i+1,j+1} = u_{i+1,j} + u_{i,j+1} + \ph(u_{i,j}) , \ee 
a trivial approximation to the class (\ref{i3}). This calculation is an example of  classification problem for a class depending on one unknown function of one variable. This class  contains trivial approximations of some well-known integrable equations included in the class (\ref{i3}), namely, the sine-Gordon, Tzitz\`eika and Liouville equations. Section \ref{conc} contains some conclusive remarks. 

\section{Preliminary definitions}\label{s2}

As eq. (\ref{i1}) has no explicit dependence on the point
$(i,j)$ of the lattice, we assume that the same will be for the generalized symmetries and conservation laws we will be considering in the following. For this reason,  without loss of generality, we  write down symmetries and conservation laws at the point $(0,0)$.
Thus eq. (\ref{i1}) can be written as: \beq\label{a1} u_{1,1} = f_{0,0} = F(u_{1,0}, u_{0,0}, u_{0,1}) . \ee
Whenever convenient we will express our formulas in terms of the two shift operators, $T_1$ and $T_2$: 
\beea \label{a1a}
T_1 u_{i,j} = u_{i+1,j}, \qquad T_2 u_{i,j} = u_{i,j+1}.
\eea
To get a scheme which is invertible and to provide propagation in both discrete directions, we have to suppose that the function $F$ depends on all its variables, i.e. 
\beq\label{a2} \pa_{u_{1,0}} F \;\cdot \pa_{u_{0,0}} F \;\cdot \pa_{u_{0,1}} F  \ne 0 . \ee

The functions $u_{i,j}$ are related among themselves by eq. (\ref{a1}) and its shifted values
\beea \nonumber
u_{i+1,j+1} = T_1^i T_2^j f_{0,0} = f_{i,j} = F(u_{i+1,j}, u_{i,j}, u_{i,j+1} ),
\eea
and it is easy to see that all of them can be expressed in terms  of the functions
 \beq\label{a3} u_{i,0} , \quad u_{0,j} , \ee where $i,j$ are arbitrary integers. This is not the only possible choice of independent variables \cite{av}, but, being the simplest, is the one we will use in the following. The functions (\ref{a3}) play the role of boundary--initial conditions for eq. (\ref{a1}).

 The evolutionary form of a generalized symmetry of eq.
(\ref{a1}) is given by the following equation
\beq\label{a4}
\frac{d}{dt} u_{0,0} = g_{0,0} = G(u_{n,0}, u_{n-1,0}, \dots,
u_{n',0}, u_{0,k}, u_{0,k-1}, \dots, u_{0,k'}) ,
\ee
where $n \ge n'$, $k \ge k'$. The form of this equation at the various points of the lattice
is obtained by applying the shift operators $T_1$ and $T_2$: 
\[ \frac{d}{dt} u_{i,j} = T_1^i T_2^j g_{0,0} = g_{i,j} = G(u_{i+n,j}, \dots, u_{i+n',j}, u_{i,j+k}, \dots, u_{i,j+k'}) . \]
Eq. (\ref{a4}) is a generalized symmetry of eq. (\ref{a1}) if the two equations (\ref{a1}, \ref{a4}) are compatible for all independent variables (\ref{a3}), i.e. 
\beea \label{a4a}
\frac{d u_{1,1}}{dt} - \frac{d f_{0,0}}{dt} \biggl|_{u_{1,1} = f_{0,0}} = 0 .
\eea
In practice, eq. (\ref{a4a}) reads
\beq\label{a5}  g_{1,1} = ( g_{1,0} \pa_{u_{1,0}} + g_{0,0} \pa_{u_{0,0}} + g_{0,1} \pa_{u_{0,1}} ) f_{0,0} . \ee 
Eq. (\ref{a5}) must be identically satisfied when all the variables $u_{i,j}$ contained in the functions $g_{i,j}$ and in the derivatives of $f_{0,0}$ are expressed in terms
of the independent variables (\ref{a3}). This result provides strict conditions, given by a set of equations for the functions $F$ and $G$, often overdetermined.

Let us consider some autonomous functions $p_{0,0},q_{0,0}$ which depend on a finite number of  functions $u_{i,j}$ and have no explicit dependence on the point $(i,j)$ of the
lattice. The relation \beq\label{a6} (T_1 -1) p_{0,0} = (T_2 -1) q_{0,0} \ee is called a (local $i,j$-independent) conservation law of eq. (\ref{a1}) if it is satisfied on the solutions set of this equation. To check it, we  need to express all variables in terms of  the independent variables
(\ref{a3}) and require that it is identically satisfied.

Starting from the choice of the independent variables (\ref{a3}) and the class of autonomous difference equations (\ref{a1}), we can prove a few useful statements  which will be used for
studying the compatibility condition (\ref{a5}).  Let us consider the functions $u_{i,1},u_{1,j}$ appearing in eq. (\ref{a5}). We can prove the following Theorem: 

\begin{The}\label{t1} The transformation $\mathcal T: \{ u_{i,0}, u_{0,j} \} \rightarrow \{ \ti u_{i,0}, \ti u_{0,j} \}$, given by the shift operator $T_2$
\beq\label{c1} \ti
u_{0,j} = u_{0,j+1} , \qquad \ti u_{i,0} = u_{i,1} , \quad i \ne 0 , \ee is invertible under the equation (\ref{a1}). Moreover, if a function $\phi$ is non--zero, then $T_2 \phi \ne 0$ too. \end{The}

\proof The invertibility of the transformation $\ti u_{0,j} = u_{0,j+1}$ is obvious. Let us show by induction that for any $i \ge 1$
\beq\label{c2} \ti u_{i,0} = \ti u_{i,0} (u_{i,0}, u_{i-1,0}, \dots, u_{1,0}, u_{0,0}, u_{0,1}) , \quad \pa_{u_{i,0}} \ti u_{i,0} \ne 0 , \quad \pa_{u_{0,1}} \ti u_{i,0} \ne 0.\ee
It follows from eq. (\ref{a1}) and condition ( \ref{a2}) that the proposition is true for $\ti u_{1,0} = u_{1,1}$. For $i
\ge 1$, from eq. (\ref{a1}) we get  \beq\label{c3} \ti u_{i+1,0} = u_{i+1,1} = F(u_{i+1,0}, u_{i,0}, \ti u_{i,0}) , \ee with $\ti u_{i,0}$ given by eq. (\ref{c2}). So
$\ti u_{i+1,0}$ has the same structure as $\ti u_{i,0}$ and thus eq. (\ref{c2}) is true. As $\ti u_{i,0}$ depends on $u_{0,1}$, then the functions $u_{i+1,0}, u_{i,0}, \ti
u_{i,0}$ are functionally independent, i.e.  $\pa_{u_{i+1,0}} \ti u_{i+1,0} \ne 0$ and $\pa_{u_{0,1}} \ti u_{i+1,0} \ne 0$.
A similar analysis can be carried out in the case of the functions $\ti u_{i,0}$ when $i \le -1$. In this case we have \beq\label{c4} \ti u_{i,0} = \ti
u_{i,0} (u_{i,0}, u_{i+1,0}, \dots, u_{-1,0}, u_{0,0}, u_{0,1}) , \qquad \pa_{u_{i,0}} \ti u_{i,0} \ne 0 , \quad \pa_{u_{0,1}} \ti u_{i,0} \ne 0. \ee From eqs. (\ref{c2}, \ref{c4}) it
follows that the transformation (\ref{c1}) is invertible.

To prove  the second part of this Theorem, let us consider a non-constant function $\phi \ne 0$. Taking into account  eq. (\ref{a1}) and its shifted values, $\phi$ can always be expressed in terms of the independent variables as
\beq\label{c5} \phi = \Phi (u_{N,0}, u_{N-1,0}, \dots, u_{N',0}, u_{0,K}, u_{0,K-1}, \dots, u_{0,K'}) , \ee for some integer numbers $N$, $N'$, $K$ and $K'$ such that  $N \ge N'$, $K \ge K'$.
Then we will have \beq\label{c6} T_2 \phi = \Phi (\ti u_{N,0}, \dots, \ti u_{N',0}, \ti u_{0,K}, \dots, \ti u_{0,K'}) . \ee
If $\phi$  depends essentially on the variables $u_{i,0}$ with $i \ne 0$, then there must exist two numbers $N$ and $N'$ such that $\pa_{u_{N,0}} \phi \ne 0$ and $\pa_{u_{N',0}} \phi \ne 0$.
When $N>0$, from eq. (\ref{c2}) it follows that only the function $\ti u_{N,0}$ appearing in eq. (\ref{c6}) depends on $u_{N,0}$. Hence
$\pa_{u_{N,0}} T_2 \phi \ne 0$, i.e. $T_2 \phi \ne 0$. The case, when  $N'<0$, is analogous. If $\phi$ depends only on $u_{0,j}$, then $\pa_{u_{0,K}} \phi \ne 0$ and $\pa_{u_{0,K'}} \phi \ne 0$, and the proof is obvious. \qed\bigskip

The operators $T_1, T_1^{-1}, T_2^{-1}$ act on the variables (\ref{a3}) in an analogous way. Consequently they also  define invertible transformations. As a result we can state
the following Proposition:

\begin{prop}\label{t2} For any non-zero function $\phi$,  $T_1^l T_2^m \phi \ne 0$ for any $l,m \in \Zz$. \end{prop}

From eqs. (\ref{c1}, \ref{c2}, \ref{c4}) we can derive the structure  of some of the partial 
derivatives of the functions $u_{i,1}$. For convenience, from now on we will define \beq\label{c7} f_{u_{i,j}} = \pa_{u_{i,j}} f_{0,0} , \qquad g_{u_{i,j}} = \pa_{u_{i,j}}
g_{0,0} \ee for the derivatives of the functions $f_{0,0}$ and $ g_{0,0}$ appearing in  eqs. (\ref{a1}, \ref{a4}). Then, for example, from eq. (\ref{a1})  we  get $\pa_{u_{1,0}} u_{1,1} =
f_{u_{1,0}}$. For  $i>0$, from eqs. (\ref{c2}, \ref{c3}) it follows that $\pa_{u_{i+1,0}} u_{i+1,1} = T_1^i \pa_{u_{1,0}} u_{1,1}$. From eq. (\ref{a1}) we can also get $u_{-1,1} = \hat F(u_{-1,0}, u_{0,0}, u_{0,1})$ and then by differentiation 
\beq\label{c8}
\pa_{u_{-1,0}} u_{-1,1} = - T_1^{-1} \frac{f_{u_{0,0}}}{f_{u_{0,1}}} .
\ee 
Then, applying the operator $T_1^{i+1}$,  with  $i<0$, to eq. (\ref{c8}) it follows that $\pa_{u_{i,0}} u_{i,1} = - T_1^i \frac{f_{u_{0,0}}}{f_{u_{0,1}}}$. For the functions of the form $u_{1,j}$ we get similar results. So  we can state the following Proposition:

\begin{prop}\label{t3} The functions $u_{i,1},u_{1,j}$ are such that  \beq\label{c9}\bea{lll} & i>0: & u_{i,1} = u_{i,1} (u_{i,0},
u_{i-1,0}, \dots, u_{1,0}, u_{0,0}, u_{0,1}) , \quad \pa_{u_{i,0}} u_{i,1} = T_1^{i-1} f_{u_{1,0}} ; \\ & i<0: & u_{i,1} = u_{i,1} (u_{i,0},
u_{i+1,0}, \dots, u_{-1,0}, u_{0,0}, u_{0,1}) , \quad \pa_{u_{i,0}} u_{i,1} = - T_1^i \frac{f_{u_{0,0}}}{f_{u_{0,1}}} ; \\ & j>0: & u_{1,j} =
u_{1,j} (u_{1,0}, u_{0,0}, u_{0,1}, \dots, u_{0,j-1}, u_{0,j}) , \quad \pa_{u_{0,j}} u_{1,j} = T_2^{j-1} f_{u_{0,1}} ; \\ & j<0: & u_{1,j} =
u_{1,j} (u_{1,0}, u_{0,0}, u_{0,-1}, \dots, u_{0,j+1}, u_{0,j}) , \quad \pa_{u_{0,j}} u_{1,j} = - T_2^j \frac{f_{u_{0,0}}}{f_{u_{1,0}}} . \ea\ee
\end{prop}

\section{Integrable example}\label{s3a}

In this Section we show, using a simple example, that effectively there are integrable equations which possess hierarchies of generalized symmetries of the form postulated in eq. (\ref{a4}) and are not included in the ABS lists.

As it is well-known \cite{lss}, 
the modified Volterra equation \beq\label{b1} u_{i,t} = (u_i^2 -1) (u_{i+1} - u_{i-1}) \ee is transformed into the Volterra
equation $v_{i,t} = v_i (v_{i+1} - v_{i-1})$ by two discrete Miura transformations: \beq\label{b2} v_i^\pm = (u_{i+1} \pm 1) (u_i \mp 1) . \ee
For any solution $u_i$ of eq. (\ref{b1}), one obtains by the transformations (\ref{b2}) two solutions $v_i^+,v_i^-$ of the Volterra equation.
From a solution of the Volterra equation $v_i$ one obtains two solutions $u_i$ and $\ti u_i$ of the modified Volterra equation.
 The
 composition of the Miura transformations (\ref{b2}) \beq\label{b3} v_i = (u_{i+1} +1) (u_i -1) = (\ti u_{i+1} -1) (\ti u_i +1) \ee provides a B\"acklund transformation for eq. (\ref{b1}). Eq. (\ref{b3}) allows one to construct, starting with a solution $u_i$ of the modified Volterra equation  (\ref{b1}), a new solution
$\ti u_i$.

Introducing for any index $i$ $u_i = u_{i,j}$ and $\ti u_i = u_{i,j+1}$, where $j$ is a new index, we can rewrite the B\"acklund transformation
(\ref{b3}) as an equation of the form (\ref{i1}). At the point $(0,0)$ it reads: \beq\label{b4} (u_{1,0} +1) (u_{0,0} -1) = (u_{1,1} -1) (u_{0,1} +1) . \ee
Eq. (\ref{b4}) does not belong to the ABS classification, as it is not invariant under the exchange of $i$ and $j$ and does not satisfy the 3D--consistency property \cite{ly09}.
The modified Volterra equation (\ref{b1}) can then be interpreted  as a 3 points 
generalized symmetry of eq. (\ref{b4}) involving only shifts in the $i$ direction: \beq\label{b5} u_{0,0,t} = (u_{0,0}^2 -1)
(u_{1,0} - u_{-1,0}) . \ee There exists also a generalized symmetry involving only shifts in the $j$ direction, given by \beq\label{b6} u_{0,0,\tau} =
(u_{0,0}^2 -1) \(\frac1{u_{0,1}+u_{0,0}} - \frac1{u_{0,0}+u_{0,-1}}\) , \ee
which belongs, together with eq. (\ref{b5}), to the complete list of integrable Volterra type equations presented in \cite{y83,y06}. Both equations have a hierarchy of generalized symmetries which, by construction, must  be compatible with eq. (\ref{b4}). Symmetries of eq.
(\ref{b5}) can be obtained in many ways, see e.g. \cite{y06}. Symmetries of eq.
(\ref{b6}) can be constructed, using the master symmetry presented in
\cite{CY}. The simplest generalized symmetries of eqs. (\ref{b5}) and (\ref{b6}) are given by the  following equations: 
\beea \nonumber u_{0,0,t'} &=& (u_{0,0}^2 -1) ( (u_{1,0}^2 -1) (u_{2,0}+u_{0,0}) - (u_{-1,0}^2 -1) (u_{0,0}+u_{-2,0}) ) , \\ \nonumber
 u_{0,0,\tau'} &=& \frac{u_{0,0}^2 -1}{(u_{0,1}+u_{0,0})^2} \( \frac{u_{0,1}^2 -1}{u_{0,2}+u_{0,1}} + \frac{u_{0,0}^2 -1}{u_{0,0}+u_{0,-1}} \)  \\ \nonumber 
 &-& \frac{u_{0,0}^2 -1}{(u_{0,0}+u_{0,-1})^2} \( \frac{u_{0,0}^2 -1}{u_{0,1}+u_{0,0}} + \frac{u_{0,-1}^2 -1}{u_{0,-1}+u_{0,-2}} \). \eea
As it can be checked by direct calculation, these equations are 5 points symmetries of eq. (\ref{b4}).

Moreover, eq. (\ref{b4}) possesses two conservation laws (\ref{a6}) characterized by the following functions $p_{0,0},q_{0,0}$: \beea\label{b7} p_{0,0}^+ &= \log
\frac{u_{0,0}+u_{0,1}}{u_{0,0}+1} , \qquad q_{0,0}^+ &= - \log (u_{0,0}+1) , \\ \label{b8} p_{0,0}^- &= \log \frac{u_{0,0}+u_{0,1}}{u_{0,1}-1}
, \qquad q_{0,0}^- &= \log (u_{0,0}-1) . \eea  It is easy to check that  eq. (\ref{a6}) is identically satisfied on the solutions of
  eq. (\ref{b4}) when we introduce into it the functions (\ref{b7}) or (\ref{b8}).
 Eq. (\ref{b4}) possess also non--autonomous conservation laws, however, conservation laws of this kind will not be discussed here.

A more general form of both eqs. (\ref{b3}, \ref{b4}) is given by
\beq\label{f7} 
v_{i,j} = (u_{i+1,j} + \al_j) (u_{i,j} - \al_j) = (u_{i+1,j+1} - \al_{j+1}) (u_{i,j+1} + \al_{j+1}) , 
\ee
where $\al_j$ is a $j$-dependent function. For any $j$ the function $u_{i,j}$ satisfies the  modified Volterra equation 
\[ 
u_{i,j,t} = (u_{i,j}^2 - \al_j^2) (u_{i+1,j} - u_{i-1,j}) 
\]
depending on the function $\al_j$. Function $v_{i,j}$, for any $j$, is a solution of the Volterra equation. Using eq. (\ref{f7}) and starting from an initial solution $v_{i,0}$, we can construct new solutions of the Volterra equation:
\[
v_{i,0} \rightarrow u_{i,1} \rightarrow v_{i,1} \rightarrow u_{i,2} \rightarrow v_{i,2} \rightarrow \dots \ .
\]
The Lax pair for eq. (\ref{f7}) is given by  
\[
L_{i,j} = 
\left( \begin{array}{cc} 
\lambda -\lambda^{-1} &  -v_{i,j} \\
1 & 0 
\end{array} \right) ,
\]
which corresponds to the standard scalar spectral problem of the Volterra equation written in matrix form, and by
\[
A_{i,j} = \frac1{u_{i,j+1} - \al_{j+1}}
\left( \begin{array}{cc} 
(\lambda - \lambda^{-1}) (u_{i,j+1} - \al_{j+1}) & 2 \al_{j+1} (u_{i,j+1}^2 - \al_{j+1}^2) \\
-2 \al_{j+1} & (\lambda - \lambda^{-1}) (u_{i,j+1} + \al_{j+1})
\end{array} \right) .
\]
This Lax pair satisfies the Lax equation 
$A_{i+1,j}  L_{i,j} = L_{i,j+1} A_{i,j}$. By setting $\al_j = 1$ we get a Lax pair for eq. (\ref{b4}). A different Lax pair for this equation has been constructed in \cite{ly09}. 

Eq. (\ref{f7}) is a direct analog of well-known dressing chain 
\beq\label{f8} 
u_{j+1,x} + u_{j,x} = u_{j+1}^2 - u_j^2 + \al_{j+1} - \al_j 
\ee
which provides a way of constructing potentials 
$v_j = u_{j,x} - u_j^2 - \al_j$ for the Schr\"odinger spectral problem
\cite{sy,s}. The Lax pair given above is analogous to that of eq. (\ref{f8}) presented in \cite{sy}.
 
\section{Derivation of the integrability conditions}\label{s3}

In this Section, following the standard scheme of the generalized symmetry method, we derive from the compatibility condition (\ref{a5}) four   conditions necessary for the  integrability of eq. (\ref{a1}).

For a generalized symmetry (\ref{a4}) we suppose that if
$g_{0,0}$ depends on at least one variable of the form $u_{i,0}$, then $g_{u_{n,0}} \ne 0$ and $g_{u_{n',0}} \ne 0$, and the numbers $n,n'$ are called orders
of the symmetry. The same can be said about the variables $u_{0,j}$ and the corresponding numbers $k,k'$ if $g_{u_{0,k}} \ne 0$ and $g_{u_{0,k'}} \ne 0$.

\begin{The}\label{t4}
Let eq. (\ref{a1}) possess a generalized symmetry (\ref{a4}) of orders $n$, $n'$, $k$ and $k'$. 
Then the following relations must take place: 
\beq\label{c10} \mbox{If} \quad n > 0 \quad \Longrightarrow \quad (T_1^n -1) \log f_{u_{1,0}} = (1-T_2) T_1 \log g_{u_{n,0}} ;
\ee\beq\label{c11} \mbox{If} \quad n' < 0 \quad \Longrightarrow \quad (T_1^{n'} -1) \log \frac{f_{u_{0,0}}}{f_{u_{0,1}}} = (1-T_2) \log g_{u_{n',0}} ; \ee\beq\label{c12} \mbox{If} \quad k > 0 \quad \Longrightarrow \quad(T_2^k -1) \log
f_{u_{0,1}} = (1-T_1) T_2 \log g_{u_{0,k}} ; \ee\beq\label{c13} \mbox{If} \quad k' < 0 \quad \Longrightarrow \quad(T_2^{k'} -1) \log \frac{f_{u_{0,0}}}{f_{u_{1,0}}} = (1-T_1) \log g_{u_{0,k'}} .
\ee \end{The}

Before going over to the proof of this Theorem, let us clarify its meaning by noting that in the case of a three point symmetry with $g_{0,0} = G(u_{1,0}, u_{0,0}, u_{-1,0})$, for which $n>0$ and $n'<0$, one can use both relations (\ref{c10}, \ref{c11}). 

\proof Let us consider the compatibility condition (\ref{a5}) expressed in terms of the  independent variables (\ref{a3}). As $g_{0,0}$ depends on $u_{i,0}$ and $u_{0,j}$, the functions
$(g_{1,1},g_{1,0},g_{0,1})$  depend on  $(u_{i,1},u_{1,j})$, whose form is given by  Proposition \ref{t3}. Moreover, eq. (\ref{a5}) will contain $u_{i,0}$ with $n+1 \ge i \ge n'$ and $u_{0,j}$ with $k+1 \ge j \ge k'$.

If $n>0$, applying to eq. (\ref{a5}) the operator $\pa_{u_{n+1,0}}$ and using the results  (\ref{c9}) contained in Proposition \ref{t3}, we get: \[ T_1 T_2 (g_{u_{n,0}}) T_1^n
f_{u_{1,0}} = f_{u_{1,0}} T_1 g_{u_{n,0}} . \] Applying the logarithm to both sides of the previous equation, we obtain eq. (\ref{c10}). The other cases are obtained in a
similar way by  differentiating eq. (\ref{a5}) with respect to $u_{n',0}$, $u_{0,k+1}$, and $u_{0,k'}$. \qed\bigskip

Eqs. (\ref{c10}--\ref{c13}) can be expressed as a standard conservation law of the form (\ref{a6}), using the obvious well-known formulae: \[\bea{lll}
& T_l^m -1 = (T_l -1) (1 + T_l + \dots + T_l^{m-1}), & m>0, \\ & T_l^m -1 = (1 - T_l) (T_l^{-1} + T_l^{-2} + \dots + T_l^m), & m<0, \ea\] where
$l=1,2$. This means that, from the existence of a generalized symmetry, one can construct some conservation laws.

Theorem \ref{t4} provides integrability conditions, i.e. that for an integrable equation there must exist a function $g_{0,0}$ satisfying eqs. (\ref{c10}--\ref{c13}). The unknown function $g_{0,0}$ must depend on a finite number of independent variables. These integrability conditions turn out to be difficult to use for testing and classifying difference equations.

In the case of the differential-difference equations of Volterra or Toda type \cite{y06}, there are integrability conditions equivalent to eqs. (\ref{c10}--\ref{c13}). In order to check these integrability conditions one can use the formal variational derivatives \cite{y80,y06,d93,hm04}, defined as 
\[
\frac{\de^{(1)} \phi}{\de u_{0,0}} = \sum_{i=-N}^{-N'} \frac{\pa T_1^i \phi}{\pa u_{0,0}} , \qquad 
\frac{\de^{(2)} \phi}{\de u_{0,0}} = \sum_{j=-K}^{-K'} \frac{\pa T_2^j \phi}{\pa u_{0,0}} ,
\]
for $\phi$ given by eq. (\ref{c5}). Using such variational derivatives, for example the integrability conditions (\ref{c10}, \ref{c12}) are reduced to the following equations:
\beq\label{vd}
\frac{\de^{(2)}}{\de u_{0,0}} (T_1^n -1) \log f_{u_{1,0}} = 0 ,  \qquad 
\frac{\de^{(1)}}{\de u_{0,0}} (T_2^k -1) \log f_{u_{0,1}} = 0 , 
\ee
which do not involve any unknown function. This result is due to the fact that in this case all discrete variables are independent. In the completely discrete case the situation is essentially different. Some of the discrete variables are dependent and the variational derivatives must be calculated modulo the equation (\ref{i1}). So,  eqs. (\ref{vd}) will not be anymore valid. If we apply here the variational derivatives, we will get, at most, some partial results depending on the choice of the independent variables introduced.

The conservation laws (\ref{c10}--\ref{c13}) depend on the order of the symmetry. These conservation laws can be simplified under some assuptions on the structure of the Lie algebra of the generalized symmetries. If we assume that for a given equation we are able to get generalized symmetries for any value of $n$ and $k$, then we can derive order-independent conservation laws, using a trick
standard in the generalized symmetry method  \cite{y06}. This assumption implies that if, for example, we have a generalized symmetry of order $n$ then there must be also one of order $n+1$. This is a very constraining assumption which is not always verified, as we know from the continuous case \cite{msy}. Here it is used just as an example for the construction of simplified formulas. In fact such simplified formulas can be obtained assuming any difference between the orders of two generalized symmetries, and in next Section we consider an example with difference 2. 

So, in the following Theorem, we will assume that in addition to (\ref{a4}) a second generalized symmetry
\beea \label{c13a}u_{0,0,\ti t} = \ti g_{0,0}=\ti G(u_{\ti n,0}, u_{\ti n-1,0}, \dots,
u_{\ti n',0}, u_{0,\ti k}, u_{0,\ti k-1}, \dots, u_{0,\ti k'}) 
\eea
of orders  $\ti n, \ti n', \ti k, \ti k'$ will exist. With this assumption we shall obtain four conservation laws:
\beq\label{c14} (T_1 -1) p_{0,0}^{(m)} = (T_2 -1) q_{0,0}^{(m)} , \qquad m=1,2,3,4, \ee with  $p_{0,0}^{(m)}$ or $q_{0,0}^{(m)}$
expressed in terms of eq. (\ref{a1}).

\begin{The}\label{t5} Let eq. (\ref{a1}) possess two generalized symmetries (\ref{c13a}) and (\ref{a4}).  Then eq. (\ref{a1}) admits the conservation laws (\ref{c14}):
\beea\label{c15a} 
 n>0,  \ \ti n = n+1 \quad & \Longrightarrow & \quad m=1, \ \ p_{0,0}^{(1)} = \log f_{u_{1,0}} ; \\
\label{c15b}  n'<0,  \ \ti n' = n'-1 \quad & \Longrightarrow & \quad m=2, \ \ p_{0,0}^{(2)} = \log \frac{f_{u_{0,0}}}{f_{u_{0,1}}} ; \\
\label{c15c}  k>0, \ \ti k = k+1 \quad & \Longrightarrow & \quad m=3, \  \ q_{0,0}^{(3)} = \log f_{u_{0,1}} ; \\ 
\label{c15d}  k'<0, \ \ti k' = k'-1 \quad & \Longrightarrow & \quad m=4, \  \ q_{0,0}^{(4)} = \log \frac{f_{u_{0,0}}}{f_{u_{1,0}}} . \eea 
\end{The}

\proof Let us consider in detail just the case when $n>0, \ \ti n = n+1$. Due to Theorem \ref{t4} eq. (\ref{c10}) must be satisfied and consequently
\beq\label{c16} (T_1^{n+1} -1) p_{0,0}^{(1)} = (1-T_2) T_1
\log \ti g_{u_{n+1,0}} , \ee where $p_{0,0}^{(1)}$ is given by (\ref{c15a}). Applying the operator $-T_1$ to eq. (\ref{c10}) and adding the result
to eq. (\ref{c16}), we get the conservation law (\ref{c14}) with $m=1$, where $q_{0,0}^{(1)}$ is given by: \[ q_{0,0}^{(1)} = T_1^2 \log
g_{u_{n,0}} - T_1 \log \ti g_{u_{n+1,0}} . \] The other cases are proved in an analogous way. \qed\bigskip

So for eq. (\ref{a1}) we have  four necessary conditions of integrability: there must exist some functions of  finite range $q_{0,0}^{(1)},
q_{0,0}^{(2)}, p_{0,0}^{(3)}, p_{0,0}^{(4)}$ of the form (\ref{c5}) satisfying the conservation laws (\ref{c14}) with $p_{0,0}^{(1)}, p_{0,0}^{(2)},
q_{0,0}^{(3)}, q_{0,0}^{(4)}$ defined by eq. (\ref{c15a}--\ref{c15d}).

The following Theorem will precise the structure of the unknown functions $q_{0,0}^{(1)}$,
$q_{0,0}^{(2)}$, $p_{0,0}^{(3)}$, and $p_{0,0}^{(4)}$.

\begin{The}\label{t6} If the functions $q_{0,0}^{(1)}$,
$q_{0,0}^{(2)}$, $p_{0,0}^{(3)}$, and $p_{0,0}^{(4)}$ satisfy eq. (\ref{c14}), with $p_{0,0}^{(1)}$, $ p_{0,0}^{(2)}$,
$q_{0,0}^{(3)}$,  and $q_{0,0}^{(4)}$ given by eqs. (\ref{c15a}--\ref{c15d}),
and are written in the form (\ref{c5}), then $q_{0,0}^{(1)}$ and $q_{0,0}^{(2)}$ may depend only on the variables $u_{i,0}$, and $p_{0,0}^{(3)}$ and
$p_{0,0}^{(4)}$  on $u_{0,j}$. \end{The}

\proof Let us consider eq. (\ref{c14}) with $m=1$. The functions therein involved have the following form: \[\bea{l} p_{0,0}^{(1)} = P^{(1)} (u_{1,0},
u_{0,0}, u_{0,1}) , \qquad p_{1,0}^{(1)} = P^{(1)} (u_{2,0}, u_{1,0}, u_{1,1}) , \\ q_{0,0}^{(1)} = Q^{(1)} (u_{N,0}, \dots, u_{N',0}, u_{0,K},
\dots, u_{0,K'}) , \\ q_{0,1}^{(1)} = Q^{(1)} (u_{N,1}, \dots, u_{N',1}, u_{0,K+1}, \dots, u_{0,K'+1}) . \ea\] Let us consider the function $q_{0,0}^{(1)}$ and let us  study its dependence on the
variables $u_{0,j}$ with $j \ne 0$. Using Proposition \ref{t3}, we see that the functions $u_{i,1}$ in $p_{1,0}^{(1)}, q_{0,1}^{(1)}$ may depend only
on $u_{0,1}$.
If $K>0$, we differentiate eq. (\ref{c14}) with $m=1$ with respect to $u_{0,K+1}$ and get: $\pa_{u_{0,K+1}} q_{0,1}^{(1)} = T_2 \pa_{u_{0,K}}
q_{0,0}^{(1)} = 0$. Then, from Proposition \ref{t2}, it follows  that $q_{0,0}^{(1)}$ does not depend on $u_{0,K}$. If $K'<0$, let us differentiate
with respect to $u_{0,K'}$ and we get  $\pa_{u_{0,K'}} q_{0,0}^{(1)} = 0$. This shows that the function $q_{0,0}^{(1)}$ cannot depend on $u_{0,j}$
with $j \ne 0$.

The proof for the other cases is quite similar. \qed\bigskip

As we cannot use the formal variational derivative, we have to work directly with functions $q_{0,0}^{(1)},q_{0,0}^{(2)}, p_{0,0}^{(3)}, p_{0,0}^{(4)}$ which have the following structure :
\[\bea{lll} 
& q_{0,0}^{(m)} = Q^{(m)} (u_{N_m,0}, \dots, u_{N_m',0}) , & \quad m=1,2; \\ 
& p_{0,0}^{(l)} = P^{(l)} (u_{0,K_l}, \dots, u_{0,K_l'}) , & \quad l=3,4. 
\ea\]
In Section \ref{s4} we are going to limit ourselves to just 5 points symmetries. This will make the problem more definite in the sense that the numbers $N_m, N_m', K_l, K_l'$ will be specified and small. 

\section{Integrability conditions for 5 points symmetries}\label{s4}

From the definition of Lie symmetry,  we can construct a new symmetry by adding the right hand sides of two symmetries $u_{0,0,t} = g_{0,0}$ and $u_{0,0,\ti t} = \ti g_{0,0}$: $u_{0,0,\hat t} = \hat g_{0,0} = c_1 g_{0,0} + c_2 \ti g_{0,0}$, where $c_1,c_2$ are arbitrary constants. For example, eq. (\ref{b4})
of Section \ref{s3a} has two 3 points symmetries (\ref{b5}) and (\ref{b6}), therefore it has a 5 points generalized symmetry: \beq\label{d1} 
u_{0,0,t} = g_{0,0} = G (u_{1,0}, u_{-1,0}, u_{0,0}, u_{0,1}, u_{0,-1}) , \quad g_{u_{1,0}} g_{u_{-1,0}} g_{u_{0,1}} g_{u_{0,-1}} \ne 0 . \ee

The other known integrable examples of the form (\ref{a1})  have also 5 points generalized symmetries. We are going to use the existence of a 5 points generalized symmetry of
the form (\ref{d1}) as an \textsl{integrability criteria}. This may be a severe restriction, as there might be integrable equations with symmetries depending on more lattice points. 

In the ABS classification all 3 points generalized symmetries turn out to be Miura transformations of the Volterra equation or of the Yamilov discretization of the Krichever--Novikov equation \cite{LPSY}. If we expect to find new type integrable discrete equations of the form (\ref{a1}) these should have as generalized symmetries some new type integrable equations. One example of such equation is given by the Narita-Itoh-Bogoyavlensky \cite{na82,it87,bo88} equation 
\beq\label{f1} 
u_{0,0,t} = g_{0,0} = u_{0,0} (u_{2,0} + u_{1,0} - u_{-1,0} - u_{-2,0}) . 
\ee
We will prove in the Appendix that no equation of the form (\ref{a1}) can have eq. (\ref{f1}) as a symmetry.

We can then state the following Theorem:

\begin{The}\label{t7} If eq. (\ref{a1}, \ref{a2}) possesses a generalized symmetry of the form (\ref{d1}), then the functions
 \beq\label{d2}\bea{lll} & q_{0,0}^{(m)} = Q^{(m)} (u_{2,0}, u_{1,0}, u_{0,0}) , & \quad m=1,2; \\ & p_{0,0}^{(m)} = P^{(m)}
(u_{0,2}, u_{0,1}, u_{0,0}) , & \quad m=3,4, \ea\ee satisfy the conditions (\ref{c14}, \ref{c15a}--\ref{c15d}). \end{The}

\proof From the relations (\ref{c10}--\ref{c13}), as $n=k=1$ and $n'=k'=-1$, we are able to construct the functions: \beq\label{d3}\bea{lll} &
q_{0,0}^{(1)} = -T_1 \log g_{u_{1,0}} , & \qquad
  q_{0,0}^{(2)} =  T_1 \log g_{u_{-1,0}} , \\
& p_{0,0}^{(3)} = -T_2 \log g_{u_{0,1}} , & \qquad
  p_{0,0}^{(4)} =  T_2 \log g_{u_{0,-1}} ,
\ea\ee satisfying conditions (\ref{c14}, \ref{c15a}--\ref{c15d}). It follows from eqs. (\ref{c9}, \ref{d1}) that the function $q_{0,0}^{(1)}$ has the
structure: \[ q_{0,0}^{(1)} = \hat Q^{(1)} (u_{2,0}, u_{1,0}, u_{0,0}, u_{1,1}, u_{1,-1}) = Q^{(1)} (u_{2,0}, u_{1,0}, u_{0,0}, u_{0,1}, u_{0,-1}) .
\] 
It analogy to Theorem \ref{t6} we get that $Q^{(1)}$ cannot depend on $u_{0,1}, u_{0,-1}$. The proof for the other functions contained in eqs. (\ref{d3}) is obtained  in the same way.
\qed\bigskip

So, for a given eq. (\ref{a1}), we check the integrability conditions (\ref{c14}, \ref{c15a}--\ref{c15d}) with the unknown functions $q_{0,0}^{(m)}$ and $p_{0,0}^{(m)}$ given in the form (\ref{d2}). If the integrability conditions are satisfied, we can construct the most general unknown functions $q_{0,0}^{(m)}$ and $p_{0,0}^{(m)}$ of the form (\ref{d2}) and then, from eqs. (\ref{d3}), build the partial derivatives of $g_{0,0}$. The partial derivatives of $g_{0,0}$ must be consistent. The consistency of eqs. (\ref{d3}) imply that the {\it additional integrability conditions} 
 \beq\label{d4} g_{u_{1,0},u_{-1,0}} = g_{u_{-1,0},u_{1,0}} , \qquad g_{u_{0,1},u_{0,-1}} = g_{u_{0,-1},u_{0,1}}  \ee 
must be satisfied. If eqs. (\ref{d4}) are satisfied, we obtain the right hand side of the symmetry (\ref{d1}) up to an arbitrary
unknown function of $u_{0,0}$ of the form $\phi(u_{0,0})$. The function $\phi$ is derived by using the compatibility condition (\ref{a5}),  the {\it final integrability condition}.

The function $g_{0,0}$, so obtained, will thus be  of the form: \beq\label{d5} g_{0,0} = \Phi (u_{1,0}, u_{0,0}, u_{-1,0}) + \Psi
(u_{0,1}, u_{0,0}, u_{0,-1}) , \ee i.e. the right hand side of any 5 points symmetry (\ref{d1}) must have the form (\ref{d5}). The same result
has been obtained by Rasin and Hydon in \cite{RH}.

All known integrable autonomous equations (\ref{a1}) have symmetries of the following two types: 
\beea \label{dd1} 
& \Psi = 0  \qquad \hbox{and} \quad & \Phi_{u_{1,0}} \Phi_{u_{-1,0}} \ne 0 ;
\\ \label{dd2} 
& \Phi = 0  \qquad \hbox{and} \quad & \Psi_{u_{0,1}} \Psi_{u_{0,-1}} \ne 0 .
\eea
Thus any symmetry of the form (\ref{d1}, \ref{d5}) is the linear combination of a symmetry (\ref{dd1}) and (\ref{dd2}). However, we cannot prove this property theoretically. 

Obviously, the scheme described in this and in the previous Sections can also be applied to the simpler symmetries (\ref{dd1}) and (\ref{dd2}).
For example, in case of a symmetry given by eqs. (\ref{d5}, \ref{dd1}), just the integrability conditions (\ref{c14}, \ref{c15a}, \ref{c15b}) must be satisfied. The first two equations of eq. (\ref{d3}) allow us to construct the partial derivatives of $g_{0,0} = \Phi$. Then we check the first of the conditions (\ref{d4}). If it is satisfied, we can  find $\Phi$ up to an arbitrary
function $\phi(u_{0,0})$, which can be specified by using the compatibility condition (\ref{a5}). 

In the case of the example considered in eq. (\ref{b4}) in Section \ref{s3a}, it is easy to check that the conditions (\ref{c14}, \ref{c15a}--\ref{c15d}) are satisfied. Moreover, using the generalized symmetries (\ref{b5}, \ref{b6}) and eqs. (\ref{d3}), we easily construct four conservation laws  which are linear combinations with shift dependent parameters of the conservation laws (\ref{b7}, \ref{b8}). 

It is worthwhile to notice that integrability conditions analogous to eqs. (\ref{c14}, \ref{c15a}--\ref{c15d}) have been derived for hyperbolic systems of the form (\ref{i4}) by Zhiber and Shabat in \cite{ZS84}.

\section{A simple classification problem}\label{s5}

Here we apply the formulae introduced before to study the class of equations: \beq\label{e1} u_{1,1} = f_{0,0} = u_{1,0} + u_{0,1} + \ph(u_{0,0}) . \ee The class of equations (\ref{e1}) depends on an
unknown function $\ph$, and we require that eq. (\ref{e1})  possess a generalized symmetry of the form (\ref{d1}). To do so it must  satisfy the integrability conditions
(\ref{c14}, \ref{c15a}--\ref{c15d}, \ref{d2}). If $\ph'' = 0$, equation (\ref{e1}) is linear, and all the integrability conditions are satisfied trivially. So we require that $\ph'' \ne 0 $.

The proof that eqs. (\ref{c14}, \ref{c15a}--\ref{c15d}, \ref{d2}) are  conservation laws  is carried out by differentiating them in such a way to reduce them to simple differential equations, a scheme introduced in 1823 by Abel  \cite{abel} (see  \cite{aczel} for a review) for solving functional 
 equations. The applications of this scheme for difference equation can be found   in \cite{lw,Hy,RH2}. In \cite{RH2} the scheme was used for finding conservation laws for known equations, i.e. when the dependence of the
functions $p_{0,0}$ and $q_{0,0}$ on the symmetries and on the equation (\ref{a1}) was unknown while the difference equation  (\ref{a1}) was given. 
In \cite{SRH} the existence of a simple conservation law is used as an integrability condition. 

Here we consider the case when either $p_{0,0}$ or $q_{0,0}$ is expressed in terms of the unknown right hand side of eq. (\ref{a1}). The conservation laws are allowed to  depend on arbitrary functions of the variables $u_{1,0}, u_{0,0}, u_{0,1}$. Moreover, as it will be shown at the end of this Section, the existence of simple conservation laws is not sufficient to prove integrability. One can have nonlinear equations of this class (\ref{e1}) with two local conservation laws but with no generalized symmetry.

Let us study the class of difference equations (\ref{e1}). For later use we can rewrite eq. (\ref{e1}) in three equivalent forms, applying to it the
operators $T_1^{-1},T_2^{-1}$: \beq\label{e2}\bea{lll} u_{-1,1} &=& u_{0,1} - u_{0,0} - \ph(u_{-1,0}) , \\ u_{1,-1} &=& u_{1,0} - u_{0,0} -
\ph(u_{0,-1}) , \\ u_{-1,-1} &=& \ph^{-1} (u_{0,0} - u_{-1,0} - u_{0,-1}) . \ea\ee 
Let us consider condition (\ref{c14}) with $m=2$. Applying the shift operators $T_1^{-1},T_2^{-1}$, we rewrite it in two equivalent forms: 
\beea \label{e3} 
p_{0,0}^{(2)} - p_{-1,0}^{(2)} &=& q_{-1,1}^{(2)} - q_{-1,0}^{(2)} , 
\\ \label{e4}
p_{0,-1}^{(2)} - p_{-1,-1}^{(2)} &=& q_{-1,0}^{(2)} -
q_{-1,-1}^{(2)} , \eea 
where $p_{0,0}^{(2)} = \log \ph'(u_{0,0})$ and $q_{0,0}^{(2)}$ is given by eq. (\ref{d2}).
Taking into account eqs. (\ref{e1}, \ref{e2}),  eqs. (\ref{e3}, \ref{e4}) can be expressed in terms of the independent variables (\ref{a3}).

Eqs. (\ref{e3}, \ref{e4}) are two functional equations for $q_{0,0}^{(2)}$. By applying the   following operators: \[ \hat {\mathcal A} = \pa_{u_{0,0}} + \pa_{u_{1,0}} + \pa_{u_{-1,0}} , \qquad \hat {\mathcal B} = \pa_{u_{0,0}} - \ph'(u_{0,0}) \pa_{u_{1,0}} -
\frac1{\ph'(u_{-1,0})} \pa_{u_{-1,0}} , \]  we reduce them to partial differential equations. Using eqs. (\ref{e2}), we can show that $\hat {\mathcal A}$ annihilates any function $\Phi (u_{1,-1}, u_{0,-1}, u_{-1,-1})$.
So,  applying $\hat {\mathcal A}$ to eq. (\ref{e4}), we get: \beq\label{e5} \hat {\mathcal A} \ q_{-1,0}^{(2)} = 0. \ee The operator $\hat {\mathcal B}$ annihilates
$q_{-1,1}^{(2)}$. Thus  applying the operator $\hat {\mathcal B}$ to eq. (\ref{e3}) we get: \[ \hat {\mathcal B} \ q_{-1,0}^{(2)} = - \hat {\mathcal B} \ (p_{0,0}^{(2)} - p_{-1,0}^{(2)}). \] If we introduce the difference operator $\hat {\mathcal C} = \hat {\mathcal A} - \hat \mathcal B$, we get  \beq\label{e6} \hat \mathcal C \ q_{-1,0}^{(2)} = \hat \mathcal B \ (p_{0,0}^{(2)} - p_{-1,0}^{(2)}) . \ee From  eqs. (\ref{e5}, \ref{e6}) we also get: \beq\label{e7} [ \hat \mathcal  A, \hat \mathcal  C] \ q_{-1,0}^{(2)} =  \hat \mathcal A  \hat \mathcal B \ (p_{0,0}^{(2)} - p_{-1,0}^{(2)}) , \ee where $[ \hat \mathcal A, \hat \mathcal C]$ is the
standard commutator of two operators.
So eqs. (\ref{e5}--\ref{e7}) can be rewritten as a partial differential system for the function $q=q_{-1,0}^{(2)}$, where, as before, by the  indexes we denote partial
derivatives and by apices derivatives with respect to the argument: 
\beq\label{e8}\bea{lll} q_{u_{0,0}} + q_{u_{1,0}} + q_{u_{-1,0}} &=& 0 ,  \\ a(u_{0,0}) q_{u_{1,0}} + b(u_{-1,0}) q_{u_{-1,0}} &=&
c(u_{0,0}) - b'(u_{-1,0}) , \\ a'(u_{0,0}) q_{u_{1,0}} + b'(u_{-1,0}) q_{u_{-1,0}} &=& c'(u_{0,0}) - b''(u_{-1,0}) . \ea\ee 
The functions $a(z)$, $b(z)$ and $c(z)$ are given by \[ a(z) =
\ph'(z) + 1 , \quad b(z) = \frac1{\ph'(z)} + 1 , \quad c(z) = \frac{\ph''(z)}{\ph'(z)} , \] where $a'(z) b'(z) c(z) \ne 0$, as  $\ph''(z) \ne 0$.

The solvability of the system (\ref{e8}) depends on the following determinant: \[ \De = \left| \bea{cc}
  a(u_{0,0})  & b(u_{-1,0})  \\
  a'(u_{0,0}) & b'(u_{-1,0})
  \ea \right| .
\] We must have $\De \ne 0$. If we have $\De = 0$, as $u_{0,0}$ and $u_{-1,0}$ are independent variables, we obtain the relations $\frac{a'(u_{0,0})}{a(u_{0,0})} =
\frac{b'(u_{-1,0})}{b(u_{-1,0})} = \nu$, where $\nu$ is a constant. These relations are in  contradiction with the condition that $\ph'' \ne 0$.

If we differentiate the system (\ref{e8}) with respect to $u_{1,0}$, we easily deduce that $q_{u_{1,0}} = \al$, where $\al$ is a constant. Then from eqs. (\ref{e8}) we obtain
two different expressions for $q_{u_{-1,0}}$:   \beq\label{e9} q_{u_{-1,0}} = \frac{d(u_{0,0}) - b'(u_{-1,0})}{b(u_{-1,0})} =
\frac{d'(u_{0,0}) - b''(u_{-1,0})}{b'(u_{-1,0})} , \qquad d(z) = c(z) - \al a(z) . \ee 
If $d' \ne 0$, differentiating eq.
(\ref{e9}) with respect to $u_{0,0}$, we get $\frac{d''(u_{0,0})}{d'(u_{0,0})} = \frac{b'(u_{-1,0})}{b(u_{-1,0})} = \sigma$, where $\sigma$ is a
constant. This result is again in contradiction with the condition $\ph'' \ne 0$. So, $d = \be$, a constant, and we get  the
following ODE for $\ph$: \beq\label{e10} \ph'' / \ph' = \al \ph' + \al + \be . \ee If $\ph$ satisfies eq. (\ref{e10}), the condition (\ref{e9}) is
satisfied, and  $q_{u_{-1,0}} = \al + \be$.

The system (\ref{e8}) provides us with another partial derivative of $q$: \[ q_{u_{0,0}} = -2\al -\be, \] from which we deduce that
\[ q =q_{-1,0}^{(2)} = \al u_{1,0} - (2\al + \be) u_{0,0} + (\al + \be) u_{-1,0} + \de , \] where $\de$ is an arbitrary constant.  The integration of eq. (\ref{e10}) gives: \[ \log \ph' (z) = \al \ph (z) + (\al + \be) z + \ga , \] 
where $\ga$ is a further constant. If we introduce  these last two equations into eq. (\ref{e3}), we get
$\be u_{0,0} + \be \ph (u_{-1,0}) = 0$, which implies $\be = 0$.

Thus, we have proved that eq. (\ref{e1}) satisfies the condition (\ref{c14}) with $m=2$ if and only if \beq\label{e11} \log \ph' (z) = \al (\ph (z) +
z) + \ga , \ee with $\al \ne 0$, as $\ph'' \ne 0$. If equation (\ref{e11}) is satisfied,
 \beq\label{e12} p_{0,0}^{(2)} = \log \ph'(u_{0,0}), \qquad
q_{0,0}^{(2)} = \al (u_{2,0} - 2 u_{1,0} + u_{0,0}) + \de , \ee and these functions define a nontrivial conservation law.

If eq. (\ref{e1}), with $\ph$  given by eq. (\ref{e11}), has a generalized symmetry of the form (\ref{d1}), the other conditions (\ref{c14}, \ref{c15a}--\ref{c15d}, \ref{d2}) must be
satisfied. From eq. (\ref{c15a}) we get that the condition (\ref{c14}) with $m=1$ becomes $(T_2 -1) q_{0,0}^{(1)} = 0$. This equation has  a trivial solution,  $q_{0,0}^{(1)}$  a constant. We now look for a nontrivial solution. From eqs. (\ref{d2}) it follows that the functions
$q_{0,0}^{(1)}$ and  $q_{0,0}^{(2)}$ depend on the same set of variables. Hence $\ti q = q_{-1,0}^{(1)}$ also satisfies eqs. (\ref{e8}),
but with zeros on the right hand side. As  $q_{u_{1,0}}$ is a constant, it follows that also $q_{0,0}^{(1)}$ must be a constant, i.e. the constant solution is the most general one.
From eqs. (\ref{d3}), we get the partial derivatives of the right hand side of the symmetry (\ref{d1}), $g_{u_{1,0}}$ and $g_{u_{-1,0}}$. It is easy to verify that the first of the conditions (\ref{d4}) is not satisfied. Consequently eq. (\ref{e1}), with $\ph$ given by eq. (\ref{e11}), has no generalized symmetry of the form (\ref{d1}). 

In Section \ref{s4} we have considered the simpler symmetries (\ref{dd1}, \ref{dd2}). Using the previous reasoning, we can prove that there is no symmetry defined by eqs. (\ref{d5}, \ref{dd1}). 
Eq. (\ref{e1}) is symmetric under the involution $u_{i,j} \rightarrow u_{j,i}$. Also the conditions (\ref{c14}) with $m=3,4$ are symmetric with respect to the conditions (\ref{c14}) with $m=1,2$. So these further conditions will provide a conservation law symmetric to the one defined by eqs. (\ref{e12}) and prove that there is no symmetry given by eqs. (\ref{d5}, \ref{dd2}). 

Let us collect the results obtained so far in the following
Theorem, where the conservation laws will be written in a simplified form, omitting inessential constants.

\begin{The}\label{t8} Eq. (\ref{e1}) satisfies the integrability conditions (\ref{c14}, \ref{c15a}--\ref{c15d}, \ref{d2}) iff $\ph$ is a solution of eq.
(\ref{e11}). Eq. (\ref{e1}), when $\ph$ is given by eq. (\ref{e11}),  has two nontrivial conservation laws: \beq\label{e14}\bea{lll} (T_1-1) (\ph (u_{0,0}) + u_{0,0}) &=&
(T_2-1) (u_{2,0} - 2 u_{1,0} + u_{0,0}) ,  \\ (T_2-1) (\ph (u_{0,0}) + u_{0,0}) &=& (T_1-1) (u_{0,2} - 2 u_{0,1} + u_{0,0}) . \ea\ee However, in this case, eq. (\ref{e1}) does not have a generalized symmetry of the form (\ref{d1}) or of the form given by eqs. (\ref{d5}, \ref{dd1}) or (\ref{d5}, \ref{dd2}).
\end{The}

Let us notice that eq. (\ref{e1}) possesses the conservation laws (\ref{e14}) for any $\ph$, not only when $\ph$ satisfies eq. (\ref{e11}). However,
 the integrability conditions are satisfied only if  $\ph$ satisfies eq. (\ref{e11}), but   no generalized symmetry of the form mentioned in Theorem
\ref{t8} exists.

\section{Conclusions}\label{conc}
In this paper we have considered the classification problem for difference equations  by asking for the existence of a generalized symmetry. In this way we have obtained the lowest order integrability conditions which turn out to be written as conservation laws. The verification of the existence of finite order conservation laws is in general a very complicated problem due to the high number of unknown involved. So we reduce ourselves to the case when we have just a 5 points symmetry. In this case we easily can find some further integrability conditions which make our problem solvable. At the end we present an example of classification when we have just an arbitrary  function of one variable. 

This research is far from complete. At the moment we are working on:

\begin{enumerate}
\item
Obtaining further integrability conditions by adding extra structures;
\item
Applying the result contained in this work for testing the integrability of some discrete equations of the class (\ref{i1}) as, for example, $Q_V$ \cite{Q5};
\item 
Classifying eqs. (\ref{i1}) in the case of an arbitrary function of two variables.
\end{enumerate}  
 
\paragraph{Acknowledgments.} R.I.Y. has been partially supported by the Russian Foundation for Basic Research (Grant numbers 08-01-00440-a and 09-01-92431-KE-a). D.L. has been partially supported by PRIN Project {\it Metodi matematici nella teoria delle onde nonlineari ed applicazioni - 2006} of the Italian Ministry of Education and Scientific Research. R.I.Y. and D.L. thank the Isaac Newton Institute for Mathematical Sciences for its hospitality during the {\it Discrete Integrable Systems} Programme.

\paragraph{APPENDIX} 
 
\begin{The}
No equation of the form (\ref{a1}, \ref{a2}) can have a generalized symmetry of the form of eq. (\ref{f1}).
\end{The}
\proof
We use conditions (\ref{c10}, \ref{c11}) with $n=2$ and $n'=-2$. Applying the operators $T_1^{-1}$ and $-T_1$, we rewrite them in the form: 
\beea \label{ff1} 
p_{1,0}^{(1)} - p_{-1,0}^{(1)} &=& \log \frac{u_{0,0}}{u_{0,1}} ,
\\ \label{ff2} 
p_{1,0}^{(2)} - p_{-1,0}^{(2)} &=& \log \frac{u_{1,1}}{u_{1,0}} ,
\eea
where $p_{0,0}^{(1)}$, $p_{0,0}^{(2)}$ are given by eqs. (\ref{c15a}, \ref{c15b}). Studying conditions (\ref{ff1}, \ref{ff2}), we will use in addition to eq. (\ref{a1}) its equivalent form: 
\[
u_{-1,1} = \hat f_{0,0} = \hat F(u_{-1,0}, u_{0,0}, u_{0,1}) .
\]

The functions $p_{0,0}^{(m)}$ have the structure $p_{0,0}^{(m)} = P^{(m)}(u_{1,0}, u_{0,0}, u_{0,1})$. Therefore $p_{-1,0}^{(m)} = P^{(m)}(u_{0,0}, u_{-1,0}, \hat f_{0,0})$ and the right hand sides of eqs. (\ref{ff1}, \ref{ff2}) do not depend on $u_{2,0}$. The functions $p_{1,0}^{(m)} = P^{(m)}(u_{2,0}, u_{1,0}, f_{0,0})$ depend on $u_{2,0}$, and from eqs. (\ref{ff1}, \ref{ff2}) we get: $\pa_{u_{2,0}} p_{1,0}^{(m)} = T_1 \pa_{u_{1,0}} p_{0,0}^{(m)} = 0$. Moreover, according to Proposition \ref{t2}, 
\beq\label{f2} 
\pa_{u_{1,0}} p_{0,0}^{(m)} = 0 , \qquad m=1,2. 
\ee

From eq. (\ref{f2}) with $m=1$ we get $f_{u_{1,0}u_{1,0}} = 0$, i.e. $f_{0,0}$ can be expressed as:  
\beq\label{f3} 
f_{0,0} = a_{0,0} u_{1,0} + b_{0,0} = A(u_{0,0}, u_{0,1}) u_{1,0} + B(u_{0,0}, u_{0,1}) ,  
\ee
where $a_{0,0} \ne 0$ due to condition (\ref{a2}). Now $p_{0,0}^{(1)} = \log a_{0,0}$ and eq. (\ref{ff1}) is rewritten as: 
\beq\label{f4} 
\frac{a_{1,0}}{a_{-1,0}} = \frac{u_{0,0}}{u_{0,1}} .
\ee
Here only the function $a_{1,0}$ depends on $u_{1,0}$, and we get:
\[
\frac{d a_{1,0}}{d u_{1,0}} = \pa_{u_{1,0}} a_{1,0} + a_{0,0} \pa_{u_{1,1}} a_{1,0} = 0 .
\]
Applying to it the shift operator $T_1^{-1}$, we get the more convenient form: 
\beq\label{f5}
\pa_{u_{0,0}} a_{0,0} + a_{-1,0} \pa_{u_{0,1}} a_{0,0} = 0 .
\ee
As $a_{-1,0} \ne 0$, only two cases are possible. The first one is when  $\pa_{u_{0,0}} a_{0,0} = \pa_{u_{0,1}} a_{0,0} = 0$, i.e. $a_{0,0}$ is a constant. This is in contradiction with eq. (\ref{f4}). So, $\pa_{u_{0,0}} a_{0,0} \ne 0$ and $ \pa_{u_{0,1}} a_{0,0} \ne 0$.

From eq. (\ref{f2}) with $m=2$ we get 
\[
\frac{f_{u_{0,0}u_{1,0}}}{f_{u_{0,0}}} - \frac{f_{u_{0,1}u_{1,0}}}{f_{u_{0,1}}} = 0 .
\]
Using this equation together with eqs. (\ref{f3}, \ref{f5}), we get: 
\[
p_{0,0}^{(2)} = \log \frac{f_{u_{0,0}u_{1,0}}}{f_{u_{0,1}u_{1,0}}} = \log \frac{\pa_{u_{0,0}} a_{0,0}}{\pa_{u_{0,1}} a_{0,0}} = \log (-a_{-1,0}) .
\]
Applying $T_1$ we can rewrite eq. (\ref{ff2}) as:
\beq\label{f6}
\frac{a_{1,0}}{a_{-1,0}} = \frac{u_{2,1}}{u_{2,0}} .
\ee
Comparing eqs. (\ref{f4}, \ref{f6}) and using eq. (\ref{f3}), we get
$u_{2,1} = \frac{u_{0,0}}{u_{0,1}} u_{2,0} = a_{1,0} u_{2,0} + b_{1,0}$. 
As $a_{1,0}$, $b_{1,0}$ do not depend on $u_{2,0}$, we obtain from here $a_{1,0} = \frac{u_{0,0}}{u_{0,1}}$. Then from eq. (\ref{f4}) we obtain $a_{-1,0} = 1$. These two last results are in contradiction, thus proving the Theorem.
\qed

 \end{document}